\documentstyle[aps,preprint]{revtex}
 \def\ket{\rangle}
\begin{document}
\title{Experimental NMR Realization of A Generalized  Quantum Search Algorithm}
\author{G. L. Long$^{1,2,3,4}$,H. Y. Yan$^{1,2}$,Y. S. Li$^{1,2}$, C. C. Tu$^{1,2}$,	
J.X.Tao$^{1,2}$,H.M.Chen$^{1}$,\\ M. L. Liu$^5$, X. Zhang$^5$, J. Luo$^5$,
 L. Xiao$^{1,2,5}$, X. Z. Zeng$^5$}
\address{$^1$Department of Physics, Tsinghua University, Beijing 100084, P.R.China\\
$^2$Key Laboratory for Quantum Information and Measurements, Ministry of Education, 
Beijing 100084, 
P.R. China\\
$^3$Institute of Theoretical Physics, Chinese Academy of Sciences, Beijing, 100080, P. R. 
China\\
$^4$Center of Atomic, Molecular and Nanosciences, Tsinghua University, Beijing 100084, P. 
R. China\\
$^5$ Laboratory of Magnetic Resonance and Atomic and Molecular Physics,
\\ Wuhan Institute of Physics and Mathematics,\\  Chinese Academy of Sciences, Wuhan 
430071, P. R. China}
\date{\today}
\maketitle

\begin{abstract}

A generalized quantum search algorithm, where phase inversions  for the marked state and 
the prepared state are replaced by $\pi/2$ phase rotations,  is realized in a 2-qubit NMR 
heteronuclear system. The quantum algorithm searches a marked state with a smaller step 
compared to standard Grover algorithm. Phase matching requirement in quantum searching is 
demonstrated by comparing it with another generalized algorithm where the two phase 
rotations are $\pi/2$ and $3\pi/2$ respectively.  Pulse sequences which include non 90 
degree pulses are given.
\end{abstract}


Grover's quantum search algorithm is one of the most important development in quantum  
computation\cite{r1}. It achieves quadratic speedup in searching a marked state in an 
unordered list over classical searching algorithms.  It has many potential applications in 
various fields of interests, for instance in deciphering the DES encryption 
code\cite{science} and algorithms that need searching. Typical examples of the algorithms 
that need searching are the Simon problem\cite{simon}, the Hamilton's circuit 
problem\cite{hamilton}, the hiddens shift problem\cite{twawley} and quantum 
counting\cite{counting}. There are several generalizations of the Grover algorithm. A 
modification of the algorithm can search for a ``chain" of $m$ marked items in 
$O(\sqrt{N/m})$ iterations\cite{r2}. In the standard Grover algorithm, the quantum 
database is built on an evenly distributed quantum superposition, and a generalization is 
made to allow the algorithm to work on a biased database where the amplitudes of the items 
in a database are not even\cite{arbi}. Searching in an arbitrary entangled superposition  
is given in Ref.\cite{carlini}

In some cases, one needs a quantum searching engine that searches an item with a smaller 
step. For instance, in the Simon algorithm\cite{simon} where ${\pi\over 2}$-phase 
rotations rather than phase inversions are used and in the case where the number of marked 
states is more than $N/4$\cite{kim}, standard Grover algorithm can not be used. In 
addition, for small $N$, state of the quantum computer may not be exactly the marked state 
during the search process, and  there is  small probability that the algorithm may fail. 
In problems that certainty of success is vital, this should be avoided. A generalized 
quantum search algorithm\cite{r3,r4} suits this purpose, where the searching step can be 
anything between that of Grover algorithm and zero. This is done by replacing the two 
phase inversions in Grover's algorithm with smaller phase rotations($\phi$ for phase 
rotation of the marked state, and $\theta$ for phase rotation of the prepared state 
$|0...0\ket$ ). It has been found that with an evenly distributed database, arbitrary 
phase rotation is not applicable\cite{r3}, and only when the two phase rotations satisfy 
the phase matching requirement $\theta=\phi$ that an efficient quantum searching algorithm 
can be constructed\cite{r4}. An approximate expression $2\sin{\theta \over 2}\;\sqrt{1 
\over {N}}$ was given for the search step\cite{r4}, where $\theta$ is the rotation angle 
of the marked state\cite{r4}. Exact expressions are given in a $SO(3) $geometric 
picture\cite{r4p}.

NMR implementation of Grover's algorithm has been realized in 2-qubit and 3-qubit 
systems\cite{r5,r6,r7}. Quantum counting has also been realized in NMR 
system\cite{rcounting}. Experimental studies  are important in
demonstrating quantum algorithms, investigating  effects of gate imperfection and  
decoherence,  and in identifying problems in building a practical quantum computer. In 
this Letter, we report the experimental realization of this phase matching quantum search 
algorithm where the phase inversions are replaced by rotations through $\pi/2$ in a 2 
qubit heteronuclear system using NMR technique. To demonstrate the effect of phase 
matching,  another  experiment where   $\theta=\pi/2$, $\phi=3\pi/2$ is also performed.  
Different from standard Grover algorithm, the pulse sequences used here contain non 90 
degree pulses, and the delay pulses need not be multiples of ${1\over 4J}$. 

Grover algorithm consists of  four steps in an iteration\cite{r2}: 
1)a phase inversion of the marked state $I_{\tau}=I-2|\tau\rangle\langle\tau|$;
2) the Walsh-Hadamard transformation W; 
3) a phase inversion of the prepared state $|0\rangle$,
$I_{0}=I-2|0\rangle\langle 0|$; and 
4) the Walsh-Hadamard transformation. The operator for one Grover iteration is $Q=-W 
I_{0}WI_{\tau}$. The steps 2-4 are combined to give the inversion about average $D$ which 
has the following matrix
\begin{eqnarray}
D_{ij}=W\;I_\tau \;W=\left\{\begin{array}{ll}{2\over N} & {\rm i\ne j}\\
                               {2\over N}-1 & {\rm i=j}\end{array}\right.
							   \end{eqnarray}

In the generalized quantum search algorithm, the phase inversions are replaced by 
arbitrary phase rotations. The corresponding operators, indicated by a ``g" in the 
superscript, are
\begin{eqnarray}
Q^g=D^g\;I_\tau^g=W\;I^g_0\;W\;I^g_\tau,
\end{eqnarray}
where
\begin{eqnarray}
I_{\tau}^{g}&=&I-(1-e^{i\phi}|\tau\rangle\langle\tau|),\nonumber\\
I_{0}^{g}&=&I-(1-e^{i\theta}|0\rangle\langle 0|),\nonumber\\
D^g&=&W\;I^g_{0}\; W.
\end{eqnarray}
When $\theta=\phi=\pi$ the Grover algorithm is recovered. It is helpful to give the 
detailed expressions for a 2-qubit system. We assume that the marked state is $\tau=3$. 
The operators are,
\begin{eqnarray}
I_{\tau=3}^{g}=\left[\begin{array}{rrrr}1&0&0&0\\ 0&1&0&0\\ 0&0&1&0\\ 0&0&0&e^{i\phi}
           \end{array}
    \right],
\end{eqnarray}
and
\begin{eqnarray}
Q^{g}= W\; I_{0}^{g}\; W
 =\left[
      \begin{array}{rrrr}1&1&1&1\\ 
                         1&-1&1&-1\\
		     			 1&1&-1&-1\\
						 1&-1&-1&1\\
				         \end{array}
						   \right]
                           \left[\begin{array}{rrrr}
						   e^{i\theta}&0&0&0\\ 
						   0&1&0&0          \\ 
						   0&0&1&0 
						   \\ 0&0&0&1
						          \end{array}
						   \right]
						   \left[\begin{array}{rrrr}1&1&1&1\\ 
                                                     1&-1&1&-1\\
													 1&1&-1&-1\\
													 1&-1&-1&1\\
					              \end{array}
						   \right].
\end{eqnarray}

For demonstration in this experiment, we choose $\theta=\phi=\pi /2$. In table \ref{t1}, 
we give the state vector for the quantum computer in each step during a quantum searching 
process where $|c\ket=1/\sqrt{N}\sum_{i\ne \tau} |i\ket$. When phase matching requirement 
is not satisfied, the quantum algorithm will not work. To demonstrate this, we also give 
the state vectors of the quantum computer for the case with $\theta=\pi/2$ and 
$\phi=3\pi/2$. Both algorithms are performed for 10 searching iterations. The probability 
for finding the marked state is the square of the coefficient of $|\tau\ket$.  These state 
vectors are converted to density matrices for comparisons with experiment.

In the experiment, the working media is $H_{2}PO_{3}$.  The 2 qubits are the nuclear spins 
of the H-atom and the P-atom. The observed J-coupling between $^{1}H$
and $^{31}P$ is {\bf 647.451 Hz}. The experiment was performed on a Bruker 500 ARX NMR 
spectrometer. The frequencies for $ ^{1}H$(qubit A) and $^{31}$P (qubit B) are   500MHz
and 220MHz respectively.  The Hamiltonian for this system can be modelled 
as a two-spin system with a weak coupling interaction,
\begin{eqnarray}
H=\omega_{A}\;I_{AZ}+\omega_{B}\;I_{BZ}+2\pi J_{AB}\;I_{AZ}\;I_{BZ},
\end{eqnarray}
where $I_{AZ}=\frac {1}{2}\sigma _{ZA}$ is the angular momentum operator in the 
$\widehat{{\bf e}_{z}}$
direction for spin A. $\omega _{A}$ and $\omega _{B}$ describe the free precession 
frequencies of the nuclear spins $A$ and $B$ respectively. The magnetic field is in 
$-$$\widehat{{\bf e}_{z}}$ direction. The quantum gate operations needed in quantum 
computation can be constructed by a combination of some hard pulses and the delay pulses. 
Compared to the pulse sequences in the Grover algorithm for 2-qubit system, one needs only 
a small modification in the  pulse sequences. First let's denote the following operators
\begin{eqnarray}
X^{\theta}&=&exp[-i\theta I_{x}],\nonumber\\
\overline{{X}^{\theta}}&=&  exp[i\theta I_{x}], \nonumber\\
Y^{\theta}&=& exp[-i\theta I_{y}],\nonumber\\ 
\overline{{Y}^{\theta}}&=& exp[i\theta I_{y}],
\end{eqnarray} 
which are radio frequency(rf) pulses for rotations about $\hat{x}$-axis through $\theta$, 
$-\theta$, rotations about $\hat{y}$-axis through $\theta$ and $-\theta$ respectively.  In 
addition to these hard pulses, we also have
\begin{eqnarray}
\tau ^{t}=exp[-2\pi i\;J_{AB}\;I_{ZA}\;I_{ZB}\;t],\nonumber
\end{eqnarray}
which is a delay pulse  where the system undergoes an evolution during period $t$ in the 
doubly rotating frame. We denote $|\uparrow\rangle=|0\rangle$ and 
$|\downarrow\rangle=|1\rangle$.

We used temporal averaging\cite{r8} to produce the effective pure state $|00\rangle$. The 
pulse sequences for the temporal averaging and the Hadmard-Walsh transformation  are 
standard\cite{r5,fuliping}:
\begin{eqnarray}
P_{0}: &&I(none);\nonumber\\
P1: &&Y_{B}^{\frac {\pi}{2}}\tau ^{\frac {1}{2J}}
	X_{B}^{\frac {\pi}{2}}Y_{A}^{\frac 	{\pi}{2}}\tau ^{\frac {1}{2J}}X_{A}^{\frac 
{\pi}{2}},\nonumber\\
P2:&&Y_{A}^{\frac {\pi}{2}}\tau ^{\frac {1}{2J}}
	X_{B}^{\frac {\pi}{2}}X_{A}^{\frac 	{\pi}{2}}\tau ^{\frac {1}{2J}}X_{B}^{\frac 
{\pi}{2}},\nonumber \\
W=&&(X_{A}^{{\frac {\pi}{2}}})^{2}\overline{Y_{A}}^{\frac {\pi}{2}}(X_{B}^{{\frac 
{\pi}{2}}})^{2}\overline{Y_{B}}^{\frac {\pi}{2}}.\nonumber
\end{eqnarray}
The pulse sequences for  the generalized quantum search algorithm are obtained by 
modifying the pulse sequences used for Grover's algorithm in  Ref.\cite{r5}:
\begin{eqnarray}
I^{g}&=&Y_{A}^{\frac {\pi}{2}}\overline{X_{A}}^{\frac {\phi}{2}}\overline{Y_{A}}^{\frac 
{\pi}{2}}Y_{B}^{\frac{\pi}{2}}\overline{X_{B}}^{\frac{\phi}{2}}\overline{Y_{B}}^{\frac{\pi
}{2}}\tau ^{\frac {2\pi-\phi}{2J\pi}},\nonumber\\
D^g&=&Y_{A}^{\frac {\pi}{2}}X_{A}^{\frac {\theta}{2}}\overline{Y_{A}}^{\frac 
{\pi}{2}}Y_{B}^{\frac{\pi}{2}}X_{B}^{\frac {\theta}{2}}\overline{Y_{B}}^{\frac 
{\pi}{2}}\tau ^{\frac{2\pi-\theta}{2J\pi}}.
\end{eqnarray}
The pulse sequence for a complete generalized search  iteration $Q^g=D^g\;I^g$ is
\begin{eqnarray}
Q^g=\overline{X_{A}}^{\frac {\theta}{2}}Y_{A}^{\frac {\pi}{2}}X_{B}^{\frac 
{\theta}{2}}Y_{B}^{\frac {\pi}{2}}\tau ^{\frac{2\pi-\theta}{2J\pi}}X_{A}^{\frac 
{\phi}{2}}\overline{Y_{A}}^{\frac {\pi}{2}}X_{B}^{\frac {\phi}{2}}\overline{Y_{B}}^{\frac 
{\pi}{2}}\tau ^{\frac {2\pi-\phi}{2J\pi}}.
\end{eqnarray}

It is interesting to study the length of time in a given quantum search algorithm with  
arbitrary phases. A hard pulse takes microsecond to complete while a delay pulse takes 
about a millisecond to complete. An algorithm with large $\theta$ or $\phi$ takes less 
time to complete. Therefore in practice, it is better to use large phase rotations to make 
the computation time short if one is given the freedom. It also worth pointing that in 
general, the hard pulses may not be the multiples of ${\pi\over 2}$ as in other NMR 
quantum computations, and the delay pulses may also takes noninteger multiples of ${1\over 
4J}$. This is different from the pulse sequences used in standard Grover algorithm, for 
instance in \cite{r5,r7}.

Two sets of experiments: phase matched searching  with $\theta=\phi={\frac{\pi}{2}}$ and 
phase mismatched searching with
$\theta={\frac{\pi}{2}},\phi={\frac{3\pi}{2}}$, have been performed.
State tomography is used  to obtain the density matrices\cite{r9}. Density matrices are 
experimentally constructed for all the 10 iterations in the two sets of experiments. It 
took quite some time and labor to get them. 
  In table \ref{t2}, we have given the relative error defined as 
$\delta\rho=\frac{||\rho_{th}-\rho_{exp}||_2}{||\rho||_2}$. However, this  error is not 
solely the ''genuine" errors\cite{r10,r10p} from gate imperfection and decoherence that 
occurs during the quantum computation process. Part of the  error is caused by doing the 
integration of areas of the spectrum by hand during the density matrix construction.  It 
is interesting to note that the relative errors at later stages are sometimes even smaller 
than the early stages. For instance at step 7 in the phase-matched case, the relative 
error is only 15\% , the smallest among the 10 steps.  Similar result is observed in other 
NMR quantum computing experiment, for instance in \cite{r5}. This is perhaps because 
imperfect gate errors cancel out each other. 
For the economy of the paper length, we give only the density matrix for the 6th iteration 
where the success rate is the largest in Fig.1 for phase matched searching. In Fig.2 the 
same is plotted for the phase mismatched searching. For clarity, we plot separately the 
real and imaginary parts.
From these figures, it is seen that the agreement between theoretical prediction and 
experimental data is good. In particular, when phase matching is satisfied, the algorithm 
can find the marked state with high probability. However, when the phase matching 
requirement is seriously violated,   the probability of finding the marked state is  very 
low. Phase mismatching leads reduction in success rate, and it should be avoided in 
practice. However, Grover's algorithm has some intrinsic robustness against errors, a 
small amount of phase mismatching, like those errors from NMR pulse manipulations,  will 
not cause a big loss in the probability for finding the marked state.

In summary, a generalized quantum search algorithm has been demonstrated in a 2-qubit NMR 
system. Pulse sequences are given. Non 90 degree pulses are used and tested to give good 
performance. It also demonstrate that phase matching in quantum searching is  important.

This work is support in part by China National Science Foundation under Grant No.60073009, 
the excellent university teacher's fund of China Education Ministry, Fok Ying-Tung 
education foundation. 
\begin{table}
\begin{center}
\caption{Theoretical State Vector of the Register in Each Search Iteration}
\label{t1}
\begin{tabular}{|c|c|c|}\hline
      &phase matching&phase mismatching\\ 
      &$\theta=\phi=\pi/2$&$\theta=\pi/2$,$\phi=3\pi/2$\\ \hline     
step1:&$|\psi_{1}\rangle=0.90|\tau\rangle+0.43|c\rangle$&
$|\psi_{1}\rangle=0.25|\tau\rangle+0.97|c\rangle$\\ \hline
step2:&$|\psi_{2}\rangle=0.97|\tau\rangle+0.22|c\rangle$&
$|\psi_{2}\rangle=0.62|\tau\rangle+0.78|c\rangle$\\ \hline
step3:&$|\psi_{3}\rangle=0.65|\tau\rangle+0.76|c\rangle$&
$|\psi_{3}\rangle=0.06|\tau\rangle+1.00|c\rangle$\\ \hline
step4:&$|\psi_{4}\rangle=0.39|\tau\rangle+0.92|c\rangle$&
$|\psi_{4}\rangle=0.59|\tau\rangle+0.80|c\rangle$\\ \hline
step5:&$|\psi_{5}\rangle=0.78|\tau\rangle+0.62|c\rangle$&
$|\psi_{5}\rangle=0.36|\tau\rangle+0.93|c\rangle$\\ \hline
step6:&$|\psi_{6}\rangle=1.00|\tau\rangle+0.01|c\rangle$&
$|\psi_{6}\rangle=0.41|\tau\rangle+0.91|c\rangle$\\ \hline
step7:&$|\psi_{7}\rangle=0.80|\tau\rangle+0.60|c\rangle$&
$|\psi_{7}\rangle=0.57|\tau\rangle+0.82|c\rangle$\\ \hline
step8:&$|\psi_{8}\rangle=0.40|\tau\rangle+0.92|c\rangle$&
$|\psi_{8}\rangle=0.13|\tau\rangle+0.99|c\rangle$\\ \hline
step9:&$|\psi_{9}\rangle=0.63|\tau\rangle+0.77|c\rangle$&
$|\psi_{9}\rangle=0.63|\tau\rangle+0.78|c\rangle$\\ \hline
step10:&$|\psi_{10}\rangle=0.97|\tau\rangle+0.24|c\rangle$&
$|\psi_{10}\rangle=0.19|\tau\rangle+0.98|c\rangle$\\ \hline
\end{tabular}
\end{center}
\end{table}
	  
\begin{table}
\begin{center}
\caption{The errors of the experiments}
\label{t2}
\begin{tabular}{ccc}
       &$\theta=\phi={\frac{\pi}{2}}$&$\theta={\frac{\pi}{2}},\phi={\frac{3\pi}{2}}$\\
  Step1&$\% 18$&$\% 17$\\
  Step2&$\% 21$&$\% 28$\\
  Step3&$\% 20$&$\% 27$\\
  Step4&$\% 21$&$\% 21$\\
  Step5&$\% 27$&$\% 20$\\
  Step6&$\% 20$&$\% 20$\\
  Step7&$\% 15$&$\% 16$\\
  Step8&$\% 24$&$\% 33$\\
  Step9&$\% 22$&$\% 30$\\
  Step10&$\% 22$&$\% 33$\\
 \end{tabular}
\end{center}
\end{table}

\noindent Figure captions\\
\noindent Fig.1 Comparisons of theoretical and experimental density matrices 
with $\theta=\phi=\pi/2$ for iteration 6. (a) and (b) are the theoretical real part and 
imaginary part of the density matrices respectively, whereas (c) and (d) are the 
corresponding experimental ones. The probability of finding the marked state is the sum  
of squares of the 11 component which are in the upper right corners of the figures.\\
\noindent Fig.2 Same as Fig.2 but for $\theta=\pi/2,\phi=3\pi/2$.\\

\begin{thebibliography}{99}
\bibitem{r1} Grover L. K.,{\it Phys. Rev. Lett.}, {\bf 79} 325 (1997)
\bibitem{science} Brassard G., {\it Science} {\bf 275}, 627 - 628 (1997).
\bibitem{simon}Simon D. R., {\it Proceedings of 35th annual Symposium on the foundations 
of computer Sciences}, 116-123; Brassard G. and Peter H\"{o}yer. {\it Proc. of Fifth 
Israeli Symposium on Theory of Computing and Systems (ISTCS'97)}, pp. 12-23, 1997; 
\bibitem{twawley}J. Twamley, {\it J. Phys.} {\bf A33} 8973, (2000). 
\bibitem{hamilton} H. Guo et al, A quantum algorithm for finding a Hamiltonian circuit,   
{\it Commun. Theor. Phys.} {\bf 35} (2001) 385-388.
\bibitem{counting} Brassard G. et al, {\it Proc. of 25th International Colloquium on 
Automata, Languages, and Programming (ICALP'98)}, {\bf  Vol.1443} of Lecture Notes in 
Computer Science, pp. 820-831, 1998. 
\bibitem{r2} Grover L. K.,{\it Phys. Rev. Lett.},{\bf 80} 4329 (1998).	
\bibitem{arbi} Biron, D., Biham, O., Biham, E., Grassl, M., Lidar, D.A.
 {\it Lecture Notes in Computer Science} vol. 1509, 140 - 147
(Springer, 1998).		
\bibitem{carlini} A. Carlini and A. Hosoya, {\it Phys. Lett.} {\bf A 280}, 114, (2001). 
Also available at quant-ph/9909089.
\bibitem{kim}Chi D. and Kim J., {\it Chaos, Solitons, and Fractals} {\bf 10} 1689 (1999).
\bibitem{r3}Long G. L.  et al,{\it Commun. Theor. Phys.}{\bf 32} 335 (1999).
\bibitem{r4}Long G. L. et al, {\it Phys. Lett.}{\bf A 262} 27 (1999). 
\bibitem{r4p} Long G. L. et al, {\it J. Phys.} {\bf A34}, 867 (2001)
\bibitem{r5}Chuang I. L. et al,{\it Phys. Rev. Lett} {\bf  80} 3408 (1998).
\bibitem{fuliping} L.P. Fu et al, {\it Chin. J. Magn. Res.}{\bf 16}, 341 (1999)
\bibitem{r6}Jones J. A. et al,{\it Nature},{\bf 393} 344 (1998).
\bibitem{r7}Vandersypen L. M. K. et al,{\it Appl. Phys. Lett.} {\bf 76} 646 (2000).
\bibitem{rcounting} J A Jones and M Mosca, {\it Phys. Rev. Lett.}{\bf 83}, 1050 (1999)
\bibitem{r8}Knill E. et al, {\it Phys. Rev} {\bf A57} 3348 (1998).
\bibitem{r9}Chuang I. L. et al,{\it Proc. R. Soc. Lond} {\bf A 454}, 447 (1998).
\bibitem{r10} Long G. L. et al, {\it Phys. Rev} {\bf A 61}, 042305 (2000). 
\bibitem{r10p} Pablo-Norman B. and Ruiz-Altaba M., {\it Phys. Rev.} {\bf A61} 012301  
(2000)
\end{thebibliography}
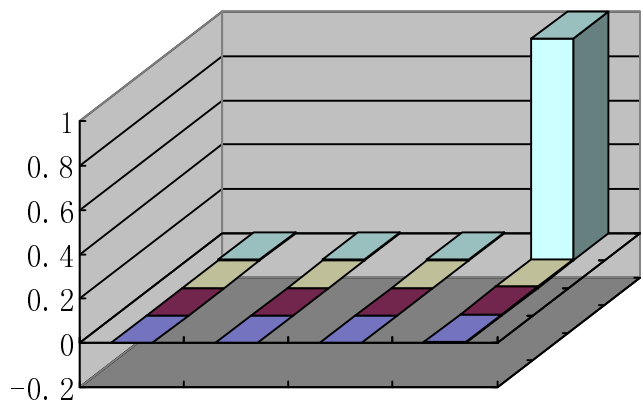
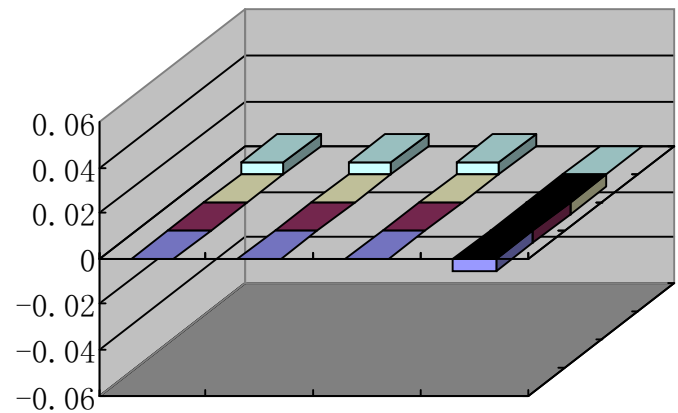
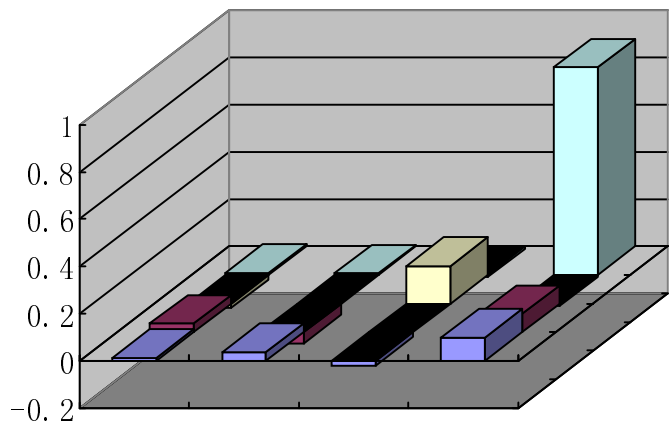
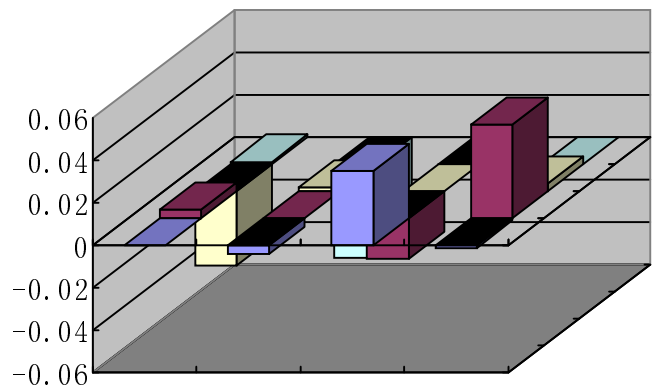
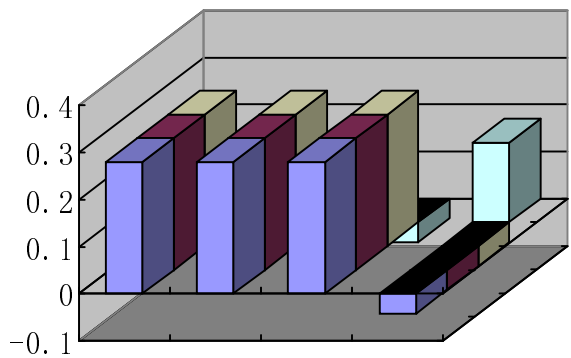
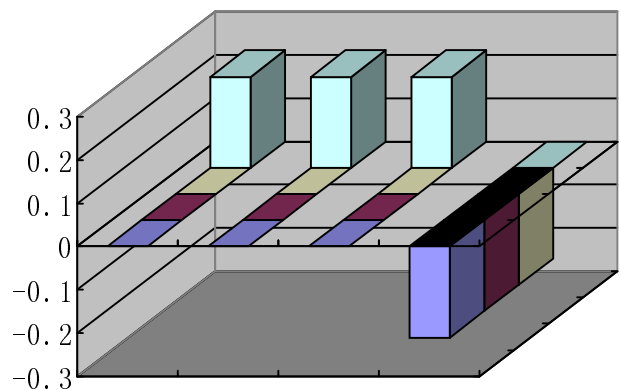
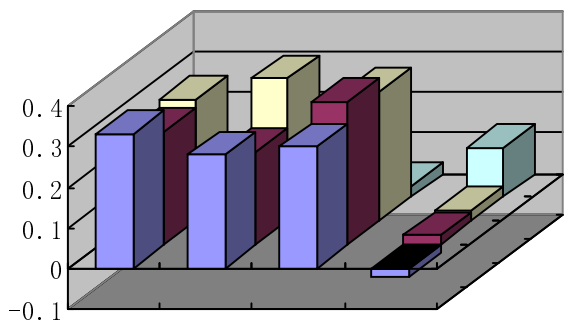
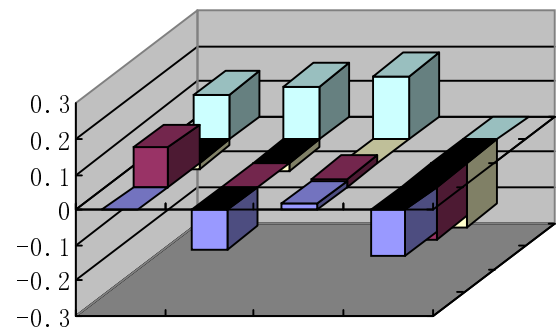
\end{document}